\documentclass[twocolumn,english,aps,prl,nofootinbib,hidelinks]{revtex4-2}
\usepackage{lmodern}
\usepackage[latin9]{inputenc}
\usepackage{geometry}
\usepackage{xcolor}
\usepackage{babel}
\usepackage{amsmath}
\usepackage{amssymb}
\usepackage{graphicx}
\usepackage[unicode=true]
 {hyperref}

\makeatletter

\DeclareFontEncoding{LGR}{}{}
\DeclareRobustCommand{\greektext}{%
  \fontencoding{LGR}\selectfont\def\encodingdefault{LGR}}
\DeclareRobustCommand{\textgreek}[1]{\leavevmode{\greektext #1}}
\ProvideTextCommand{\~}{LGR}[1]{\char126#1}

\DeclareTextSymbolDefault{\textquotedbl}{T1}
\providecommand{\tabularnewline}{\\}

\usepackage{hyperref}
\usepackage{orcidlink}

{\newpage}
\newcommand{\lb}{\linebreak}
\newcommand{\vs}{\vskip2pt}

\makeatother

\begin{document}
\title{An intrinsic kinematic relation $\boldsymbol{c=\frac{c_{0}}{H_{0}}\,\dot{a}}\,$
inferred from Type Ia supernovae}
\author{Hoang Ky Nguyen \orcidlink{0000-0003-2343-0508}$\ $}
\email[\ ]{hoang.nguyen@ubbcluj.ro}

\affiliation{{\vskip2pt}Department of Physics, Babe\c{s}-Bolyai University, Cluj-Napoca
400084, Romania\vskip1ptInstitute for Interdisciplinary Research
in Science and Education, ICISE, Quy Nhon 55121, Vietnam}
\date{\today}
\begin{abstract}
Based on the Hubble diagram of Type Ia supernovae, we present empirical
evidence for a kinematic relation between the speed of light and the
late-time cosmic expansion rate. To infer this relation, we employ
the Dolgov--Barrow cosmology, described by Dolgov's power-law expansion
$a=\left(t/t_{0}\right)^{\,\mu}$ and Barrow's varying-speed-of-light
(VSL) $c=c_{0}\,a^{-\zeta}$. In this cosmology, light propagating through
an expanding cosmic background experiences an additional refraction
induced by the variation of $c$ along its path, resulting in a \emph{modified}
Lema\^itre redshift relation $1+z=a^{-(1+\zeta)}$. The model provides
a high-quality fit to the Pantheon Hubble diagram of Type Ia supernovae
and exhibits \emph{a remarkably tight posterior degeneracy along the
locus }$(1+\zeta)\,\mu=1$. In particular, the case $\left\{\mu=2/3,\,\zeta=1/2\right\}$,
referred to as the VSL--Einstein--de Sitter case, is favored over
flat $\Lambda$CDM by $\Delta\chi^{2}=2.8$, i.e. at $68\%$ confidence
level, despite having the same number of free parameters. The empirical
relation $(1+\zeta)\,\mu=1$ entails that \emph{the speed of light
is exactly proportional to the cosmic expansion rate}, $c=c_{0}\,H_{0}^{-1}\,\dot{a}$,
throughout late-time cosmic evolution, a synchronous behavior absent
in the standard $\Lambda$CDM model. Although the empirical relation
$(1+\zeta)\,\mu=1$ is inferred using the Dolgov--Barrow parameterization,
the resulting expression $c=c_{0}\,H_{0}^{-1}\,\dot{a}$ is an \emph{intrinsic}
relation because it relates two physical quantities of the same dimension:
the speed of light and the cosmic expansion rate. It is therefore
independent of arbitrary choices of units or parameterization and
encodes a purely kinematic correspondence between $c$ and $\dot{a}$.
If confirmed by independent cosmological probes, this relation may
point toward a more general kinematic principle governing late-time
cosmic evolution. We discuss several implications of this relation
for late-time cosmology, including an alternative interpretation of
cosmic acceleration.
\end{abstract}
\maketitle
\textbf{\emph{Motivation}}---The standard cosmological paradigm,
based on the Lambda Cold Dark Matter ($\Lambda$CDM) model, has been
very successful in explaining a wide range of observations---from
the cosmic microwave background to large-scale structure formation---yet
it fundamentally relies on the existence of an enigmatic form of dark
energy (DE), represented by the cosmological constant $\Lambda$ \citep{Perivolaropoulos-2022,diValentino-2021,Bull,Freedman:2021abc}.\vs

The most \emph{direct} evidence of DE comes from the Hubble diagram
of Type Ia supernovae (SNeIa), which indicates late-time cosmic acceleration
\citep{Riess-1998,Perlmutter-1999}. The evidence for DE deduced from
the Cosmic Microwave Background (CMB) is \emph{indirect}, however.
In an inherently groundbreaking work in 2003 \citep{BDRS-2003}, Blanchard,
Douspis, Rowan-Robinson and Sarkar (BDRS) achieved an excellent fit
to the CMB power spectrum within the Einstein--de Sitter (EdS) universe,
viz.\emph{ without invoking DE}, by slightly modifying the CMB primordial
fluctuations spectrum. (We have recently elucidated BDRS's findings
in Ref.\ \citep{VSL2024-Pantheon}.) The success of BDRS's approach
raises a possibility that the late-time acceleration may also be explained
within the EdS model \emph{by modifying some other assumption in cosmology}.\vs

To illustrate, let us revisit the spatially-flat Robertson--Walker
(RW) metric (where $d\Omega_{(2)}^{2}:=d\theta^{2}+\sin^{2}\theta d\varphi^{2}$):
\begin{equation}
ds^{2}=c^{2}\,dt^{2}-a^{2}(t)\left(dr^{2}+r^{2}d\Omega_{(2)}^{2}\right)\label{eq:RW}
\end{equation}
The EdS model corresponds to the growth law $a(t)\propto t^{\,2/3}$.
It is known that this model fails to account for the high redshift
section of the Hubble diagram of SNeIa: supernovae with $z\gtrsim1$
appear dimmer than expected from the EdS model. The $\Lambda$CDM
model addresses this problem by employing $\Lambda$ to \emph{modify
}$a(t)$. Ignoring radiation, the Friedmann equation of the spatially-flat
$\Lambda$CDM model has an exact solution
\begin{equation}
a(t)=\!\Bigl(\frac{1-\Omega_{\Lambda}}{\Omega_{\Lambda}}\Bigr)^{1/3}\Bigl(\sinh\frac{t}{t_{\Lambda}}\Bigr)^{2/3}\,;\ \ \ t_{\Lambda}:=\!\frac{2}{3H_{0}\sqrt{\Omega_{\Lambda}}}\label{eq:LCDM-growth}
\end{equation}
With $\Omega_{\Lambda}\approx0.73$ and $H_{0}\approx70$ km/s/Mpc,
the $\Lambda$CDM solution produces an excellent fit to the Hubble
diagram of SNeIa. Its success has been interpreted as \emph{direct}
evidence for DE. Nevertheless, this function $a(t)$ introduces a
\emph{preferred} time scale $t_{\Lambda}\sim11$ Gy which is comparable
with the universe's current age $t_{0}\approx13.7$ Gy, given by $t_{0}\!=\!\frac{2\,\text{atanh}\sqrt{\Omega_{\Lambda}}}{3H_{0}\sqrt{\Omega_{\Lambda}}}$.
The near equality between $t_{\Lambda}$ and $t_{0}$ represents the
\emph{``why-now''} coincidence puzzle.\vs

Recently, in a series of articles \citep{VSL2024-Pantheon,VSL2024-HiggsDilaton,VSL2024-dilaton},
we explored an alternative cosmological approach in which the EdS
expansion law $a(t)\propto t^{\,2/3}$ is retained \emph{while the
speed of light $c$ varies with the cosmic scale factor $a$} in the
RW metric \eqref{eq:RW}. The concept of a variable speed of light
(VSL) dates back to Einstein's 1911 discussion of the influence of
gravity on light \citep{Einstein-1911,Einstein-1912-a,Einstein-1912-b}
and was later revived in modern cosmology by Moffat \citep{Moffat-1993}
and by Albrecht and Magueijo \citep{Albrecht-1998}. In particular,
in \citep{VSL2024-HiggsDilaton,VSL2024-dilaton}, we considered a
scale-invariant theory in which matter couples \emph{non-minimally}
to gravity through a dilaton field, leading to $a=\left(t/t_{0}\right)^{\,2/3},\ c=c_{0}\,a^{-1/2}$.\vs

Motivated by this development, in the current paper, we will approach
the late-time acceleration from a novel perspective: we will seek
to deduce \emph{empirical} information about the \emph{kinematic}
functions $a(t)$ and $c(a)$ directly from observational data. That
is to say, rather than imposing \emph{ab initio} a specific first-principles
mechanism that might be responsible for late-time acceleration---such
as the $\Lambda$CDM model or any modified gravity theory---we will
instead investigate how the Hubble diagram of SNeIa may dictate the
functions $a(t)$ and $c(a)$. This new knowledge of the \emph{kinematics}
of cosmology can subsequently guide theorists in formulating the \emph{dynamics}
of cosmology.\vs

To proceed, we need to specify the functions $a(t)$ and $c(a)$ in
parsimonious forms. To avoid the \emph{``why-now''} problem (which
the $\Lambda$CDM model encounters as mentioned above), $a(t)$ and
$c(a)$ must not involve a preferred scale. Under the assumption that
no preferred scale exists, the only admissible mathematical form is
a \emph{power law}, as proved in Appendix \ref{sec:Power-law-derivation}.
Therefore, we will consider the VSL power-law cosmology, described
by the VSL--RW metric:
\begin{align}
ds^{2} & =c^{2}(a)\,dt^{2}-a^{2}(t)\left(dr^{2}+r^{2}d\Omega^{2}\right)\label{eq:modified-RW}\\
a(t) & =\left(t/t_{0}\right)^{\mu}\label{eq:EdS-growth}\\
c(a) & =c_{0}\,a^{-\zeta}\label{eq:VSL-c-general}
\end{align}
where $c_{0}$ is the speed of light in cosmic space at the current
time $t_{0}$. This metric reflects the power-law cosmology $a\!\propto\!t^{\,\mu}$
originated by Dolgov \citep{Dolgov:1997hju} and a varying speed of
light $c\!\propto\!a^{-\zeta}$ proposed by Barrow \citep{Barrow-1998a}.
The parameter pair $\{\mu,\zeta\}$ spans the parameter space for
our VSL power-law cosmology. The case $\left\{ \mu=1,\zeta=0\right\} $
was put forth by Kolb in the form of a coasting universe model, where
$a(t)\propto t$ \citep{Kolb:1989mnb}. The scenario with $\zeta=0$
but $\mu\in\mathbb{R}$ unconstrained was examined by Dolgov et al.
\citep{Dolgov:2014ytr} and Tutusaus et al. \citep{Tutusaus:2016kjh}.
The case $\left\{ \mu=\frac{2}{3},\zeta=\frac{1}{2}\right\} $, referred
to as the VSL--EdS case, was investigated in Ref.\,\citep{VSL2024-Pantheon}.\vs\vs

\textbf{\emph{Problem statement}}---Let us define a parameter\vspace{-.1cm}
\begin{equation}
\gamma:=1/\mu-\zeta\label{eq:gamma-def}
\end{equation}
Consider another pair $\left\{ \mu',\zeta'\right\} $ such that $1/\mu'-\zeta'=\gamma$.
It can be shown that under the cosmic time coordinate transformation
\begin{equation}
t'=t_{0}'\left(t/t_{0}\right)^{\mu/\mu'}\ \ \ \text{where}\ \ \ t_{0}':=\left(\mu'/\mu\right)t_{0}\label{eq:transf}
\end{equation}
the VSL--RW metric in \eqref{eq:modified-RW} remains unchanged by
replacing $a(t)$ and $c(a)$ with\vspace{-.1cm}
\begin{align}
a(t') & =\left(t'/t'_{0}\right)^{\mu'};\ \ \ \ \ c'(a)=c_{0}\,a^{-\zeta'}\label{eq:new-a-c}
\end{align}
It should be emphasized that, at this stage, $\gamma\in\mathbb{R}$
is still a \emph{free} parameter. \emph{The special value $\gamma=1$
would be equivalent to the identity}\vspace{-.1cm}
\begin{equation}
(1+\zeta)\,\mu=1\label{eq:special-identity}
\end{equation}
\emph{When this identity holds}, Eqs.\ \eqref{eq:EdS-growth} and
\eqref{eq:VSL-c-general} yield the following relation\vspace{-.1cm}
\begin{equation}
c=\frac{c_{0}t_{0}}{\mu}\,\frac{da}{dt}\ \ \ \ \ \forall t\label{eq:key-relation}
\end{equation}
viz. \emph{the speed of light $c$ is strictly proportional to the
cosmic expansion rate $da/dt$ at all instants}. The key objective
of our paper is to establish whether the Hubble diagram of SNeIa supports
Relation \eqref{eq:key-relation}. Equivalently, we will statistically
\emph{test the null hypothesis $(1+\zeta)\,\mu=1$}, viz. Eq. \eqref{eq:special-identity},
using the Hubble diagram of SNeIa.\vs\vs

\begin{figure}[!t]
\begin{centering}
$\!\!\!$\includegraphics[width=9.2cm,height=3.65cm]{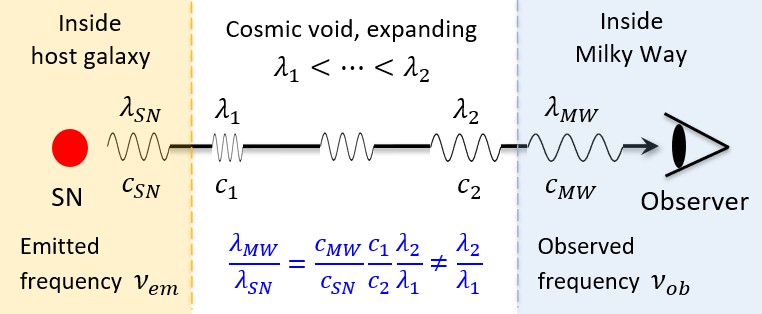}
\par\end{centering}
\caption{\label{fig:transition}The supernova emitted $\lambda_{SN}$ and the
observer detects $\lambda_{MW}$. Note that $\frac{\lambda_{MW}}{\lambda_{SN}}\protect\neq\frac{\lambda_{2}}{\lambda_{1}}$
in general, rendering the Lema\^itre redshift formula, $1+z=a^{-1}$,
\emph{invalid} for VSL cosmology.}
\end{figure}

\textbf{\emph{Modifying the Lema\^itre redshift relation}}---As
derived in Appendix \ref{sec:Freq-ratio}, the relation $c\propto a^{-\zeta}$
in Eq.\ \eqref{eq:VSL-c-general} yields the ratio between the frequencies
of the emitted and the observed lightwaves (where $a_{ob}=1$ and
$a_{em}:=a$)
\begin{equation}
\frac{\nu_{ob}}{\nu_{em}}=\frac{a_{em}^{1+\zeta}}{a_{ob}^{1+\zeta}}=a^{1+\zeta}\label{eq:freq_shift}
\end{equation}
However, converting this frequency ratio to the redshift requires
special care. In Ref.\ \citep{VSL2024-Pantheon} we clarified two
crucial requirements. Firstly, as demonstrated in Fig.\ \ref{fig:transition},
when the emitted lightwave of frequency $\nu_{em}$ exited the galaxy
hosting the supernova (SN), the difference between $c_{SN}$ and $c_{1}$
caused it to undergo a refraction and changed its wavelength from
$\lambda_{SN}$ to $\lambda_{1}$, whereby $\nu_{em}$$\,=c_{SN}/\lambda_{SN}$$=c_{1}/\lambda_{1}$.
Similarly, when the incoming lightwave of frequency $\nu_{ob}$ enters
the Milky Way to reach the Earth-based astronomer, the difference
between $c_{2}$ and $c_{MW}$ causes another refraction and changes
its wavelength from $\lambda_{2}$ to $\lambda_{MW}$, whereby $\nu_{ob}=\!c_{2}/\lambda_{2}\!=\!c_{MW}/\lambda_{MW}$.
In combination, these relations lead to
\begin{equation}
\frac{\lambda_{MW}}{\lambda_{SN}}=\frac{\nu_{ob}^{-1}\,c_{MW}}{\nu_{em}^{-1}\ c_{SN}\;}=a^{-(1+\zeta)}\,\frac{c_{MW}}{c_{SN}}\label{eq:wavelength_ratio}
\end{equation}

Secondly, the microscopic physics within the Milky Way \emph{in principle}
may differ from that in the (distant) host galaxy. In Ref.\ \citep{VSL2024-Pantheon},
we introduced the notion of a ``yardstick'' wavelength $\lambda_{MW}^{*}$,
\emph{against which the astronomer compares the observed wavelength
$\lambda_{MW}$} (rather than comparing \emph{$\lambda_{MW}$} directly
to the emitted wavelength $\lambda_{SN}$). Building on the discussion
in Ref.\ \citep{VSL2024-Pantheon}, we will adopt the relationships
$c_{MW}\propto(\lambda_{MW}^{*})^{-\zeta}$ and $c_{SN}\propto\lambda_{SN}^{-\zeta}$
(see Appendix \ref{sec:Power-law-derivation}). Along with Eq. \eqref{eq:wavelength_ratio},
they yield the redshift
\begin{equation}
1+z=\frac{\lambda_{MW}}{\lambda_{MW}^{*}}=a^{-(1+\zeta)}\biggl(\frac{\lambda_{SN}}{\lambda_{MW}^{*}}\biggr)^{1+\zeta}\label{eq:tmp}
\end{equation}
For an ensemble of galaxies at a given redshift $z$, the ratio $F:=\frac{\lambda_{SN}}{\lambda_{MW}^{*}}$
can be taken to be a function of $z$ alone. We thus arrive at the
\emph{VSL Lema\^itre redshift formula} \citep{_comm-Hubble}: 
\begin{equation}
1+z=a^{-(1+\zeta)}\,F^{1+\zeta}(z)\label{eq:VSL-Lemaitre}
\end{equation}
which fundamentally deviates from the standard Lema\^itre redshift
formula $1+z=a^{-1}$. If $F(z)\equiv1$ for all $z$, Eq.\ \eqref{eq:VSL-Lemaitre}
reduces to $1+z=a^{-(1+\zeta)}$. In Eq. \eqref{eq:VSL-Lemaitre},
the function $F(z)$ plays the role of a ``redshift modulator''. Alternative
phenomenological modifications to the redshift relation, such as the
``redshift remapping'' approach, have also been explored in \citep{Bassett:2013hjg,Wojtak:2016hfg}.\vs

Adopting the practice in Ref.\,\citep{VSL2024-Pantheon}, we will
model $F(z)$ as
\begin{equation}
F(z)=1-\delta\left[1-(1+z)^{-2}\right]^{2}\label{eq:F-param}
\end{equation}
where $\delta$ is a dimensionless parameter. The function $F(z)$
monotonically interpolates between $F(z=0)=1$ and $F(z=\infty)=1-\delta$,
\vs\vs

\textbf{\emph{Modifying the luminosity distance--redshift formula}}---In
Appendix \ref{sec:Modified-d-z}, based on Eqs.\ \eqref{eq:EdS-growth}
and \eqref{eq:VSL-Lemaitre}, we derive the \emph{modified }luminosity
distance--redshift relation:\vspace{-.1cm}
\begin{align}
d_{L}^{MW} & =\frac{c_{MW}\,t_{0}}{1-\eta}\,\frac{1+z}{F(z)}\left[1-\left(\frac{1+z}{F^{1+\zeta}(z)}\right)^{1-\frac{1}{\eta}}\right]\vspace{-.05cm}\label{eq:VSL-d-vs-z}\\
\eta\,\  & \!\!:=(1+\zeta)\,\mu\vspace{-.05cm}
\end{align}
where $d_{L}^{MW}$ is the luminosity distance observed by the Earth-based
astronomer and $c_{MW}\!=\!300,000\,$km/s is the speed of light in
the Milky Way. We note that, besides $\zeta$, Eq. \eqref{eq:VSL-d-vs-z}
depends on $\eta$, but not on the parameter $\gamma$ defined in
Eq.\ \eqref{eq:gamma-def}. The parameters are related as\vspace{-.1cm}
\begin{equation}
1-\eta=(\gamma-1)\,\mu\vspace{-.1cm}
\end{equation}
The \emph{special} value $\gamma=1$ renders $\eta=1$, or $(1+\zeta)\,\mu=1$.\vs

For a given pair of $\{\mu,\zeta\}$, Eqs. \eqref{eq:VSL-d-vs-z}
and \eqref{eq:F-param} involve two free parameters, $t_{0}$ and
$\delta$.\vs\vs

\textbf{\emph{Fitting to the Combined Pantheon Sample of SNeIa}}---In
Ref.\ \citep{Scolnic-2018} Scolnic et al. compiled the Pantheon
Catalog consisting of 1,048 SNeIa spanning the redshift range $z\in[0.01,2.26]$.
The Catalog \citep{Pantheon-data} provides for each supernova the
measured redshift $z_{i}$, apparent magnitude $m_{i}^{\text{Pan}}$
, and associated uncertainty $\sigma_{i}^{\text{Pan}}$. Throughout
this work, we adopt the absolute magnitude $M=-19.35$ to compute
the observed distance modulus $\mu_{i}^{\text{Pan}}\!:=\!m_{i}^{\text{Pan}}-\!M$,
which is related to the luminosity distance via $\mu=5\log_{10}(d_{L}/\text{Mpc})+25$.
For every point $\left\{ \mu,\,\zeta\right\} $ in the parameter space
of the model specified in Eqs. (\ref{eq:modified-RW}--\ref{eq:VSL-c-general}),
the parameters $t_{0}$ and $\delta$ are optimized by minimizing
the chi-square statistic ${\displaystyle \chi^{2}=\sum_{i=1}^{1,048}}\!\left(\!\frac{\mu_{i}^{\text{model}}-\mu_{i}^{\text{Pan}}}{\sigma_{i}^{\text{Pan}}}\!\right)^{2}$,
where $d_{L}$ is given by Eqs.\,\eqref{eq:VSL-d-vs-z} and \eqref{eq:F-param}.\vs

For comparison, we also fit the flat $\Lambda$CDM model using the
standard luminosity-distance relation $d_{L}^{\Lambda\text{CDM}}\!=\frac{c}{H_{0}}\!(1\!+\!z)\lb\int_{0}^{z}\!\!\frac{dz'}{\sqrt{(1-\Omega_{\Lambda})(1+z')^{3}+\Omega_{\Lambda}}}$,
which yields best-fit values $H_{0}\!=\!70.21$ and $\Omega_{\Lambda}\!=\!0.715$,
listed in the final row of Table I.\vs

\begin{figure}[!t]
\begin{centering}
\hskip-16pt\includegraphics[width=6.9cm,height=7.5cm]{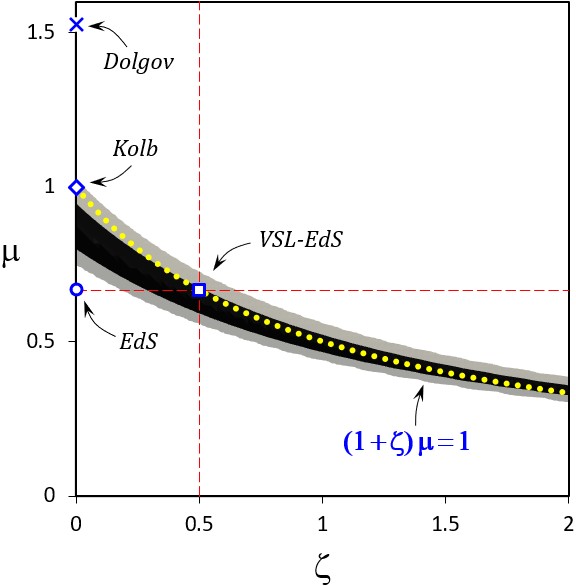}
\par\end{centering}
\noindent \centering{}\caption{\label{fig:mu-zeta}68\% and 95\% CL contours, deduced from Pantheon
Catalog and VSL power-law cosmology $\{a\!\propto\!t^{\,\mu},\,c\!\propto\!a^{-\zeta}\}$.\lb\emph{
A degeneracy tracking the locus $(1+\zeta)\,\mu\!=\!1$ is prominent.}
In Table I below, the Kolb and VSL--EdS cases outperform flat $\Lambda$CDM
in terms of $\chi^{2}$. Additional advantages of VSL--EdS over flat
$\Lambda$CDM are provided in the main text.\vspace{0.6cm}}
\begin{tabular}{cccc|ccc}
\textbf{\small{}\!Table I} & $\mu$ & $\zeta$ & $\delta$ & $\ $$\chi^{2}$ & $\!H_{0}${\footnotesize{}$\,$}{\scriptsize{}(km/s/Mpc)} & $t_{0}${\scriptsize{}$\,$(Gy)$\!\!$}\tabularnewline
\hline 
\hline 
\textbf{\!}EdS & 2/3 & 0 & 0 & $\ \:$2482.7 & 62.87 & 10.38\tabularnewline
\textbf{\!}Dolgov & 1.52 & 0 & 0 & $\ \:$1067.1 & 69.32 & 21.46\tabularnewline
\!Kolb & 1 & 0 & 0.089$\ \,$ & $\ \:$1035.0 & 70.54 & 13.87\tabularnewline
\textbf{\small{}\!VSL-EdS} & \textbf{\small{}2/3} & \textbf{\small{}1/2$\ $} & \textbf{\small{}0.069$\ \,$} & \textbf{\small{}$\ \:$1032.9} & \textbf{\small{}47.22} & \textbf{\small{}13.82}\tabularnewline
\hline 
\hline 
{\small{}\!Flat }$\Lambda$CDM & \multicolumn{3}{c|}{{\small{}$(\Omega_{\Lambda}=0.715)$}} & $\ \:$1035.7 & 70.21 & 13.63\tabularnewline
\end{tabular}\vspace{0cm}
\end{figure}

Figure \ref{fig:mu-zeta} presents the resulting confidence contours
in the $\left\{ \mu,\,\zeta\right\} $ plane. The black region denotes
the 68\% confidence level (CL), while the gray regions correspond
to the 95\% CL. Four representative cosmological models are highlighted
in the figure and summarized in Table I.\vs\vs

\paragraph{(i) Einstein--de Sitter:}

The EdS universe, $a \propto t^{2/3}$, provides a poor fit to the SNeIa Hubble diagram,
as expected. This well-known failure motivated the introduction of
a cosmological constant, leading to the flat $\Lambda$CDM model.\vs

\paragraph{(ii) Dolgov:}

The authors of Refs. \citep{Dolgov:2014ytr,Tutusaus:2016kjh} investigated
power-law cosmology $a\propto t^{\,\mu}$ \citep{Dolgov:1997hju},
while maintaining a constant speed of light (i.e. $\zeta=0$). Their
analysis did not include the redshift modulation function $F(z)$
(equivalently, $\delta=0$), yielding an optimal value $\mu\approx1.52$.
Although the Dolgov model substantially improves upon the EdS cosmology,
it remains severely non-competitive with the flat $\Lambda$CDM model,
$\chi_{\text{Dolgov}}^{2}-\chi_{\Lambda\text{CDM}}^{2}=34.2$.\vs

\paragraph{(iii) Kolb:}

The Kolb model \citep{Kolb:1989mnb} corresponds to the coasting universe
$a\propto t$, represented by the point $\left\{ \mu=1\right.$, $\left.\zeta=0\right\} $
in Fig. \ref{fig:mu-zeta}. It occupies a distinguished location within
the $\left\{ \mu,\,\zeta\right\} $ parameter space as the constant-$c$
member of the family satisfying the identity $(1+\zeta)\,\mu=1$.\vs

\paragraph{(iv) VSL--EdS:}

The VSL--EdS model, $\left\{ a\propto t^{\,2/3}\right.$, $\left.c\propto a^{-1/2}\right\} $,
was investigated in Refs.\ \citep{VSL2024-Pantheon}. This model
arises from a scale-invariant gravitational action in which matter
couples non-minimally to gravity through a dilaton field, without
introducing a dark-energy term \citep{VSL2024-dilaton,VSL2024-HiggsDilaton}.
Remarkably, the Pantheon Catalog favors this model over flat $\Lambda$CDM,
with $\chi_{\text{\ensuremath{\Lambda}CDM}}^{2}-\chi_{\text{VSL-EdS}}^{2}=2.8$.
Since both models have two free parameters, this corresponds approximately
to the 68\% confidence level.\vs
\begin{figure}[!t]
\begin{centering}
\hskip-16pt\includegraphics[width=6.55cm,height=7.5cm]{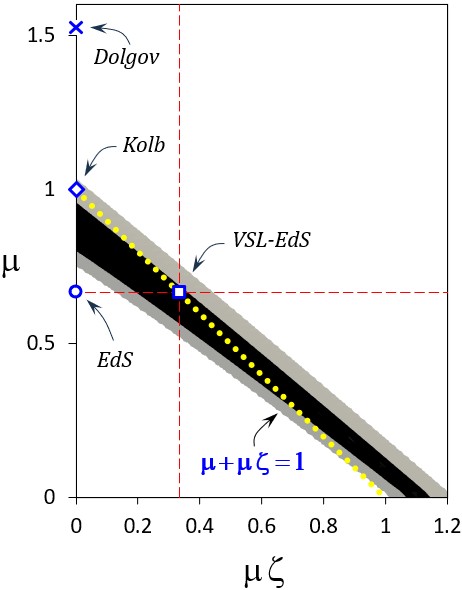}\vspace{.4cm}
\par\end{centering}
\begin{centering}
\hskip-11pt\includegraphics[width=6.3cm,height=3.8cm]{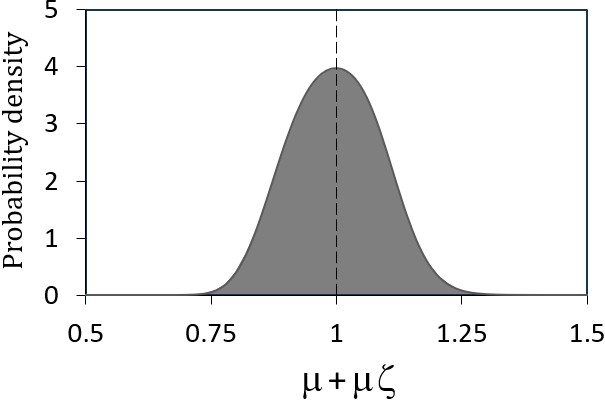}
\par\end{centering}
\noindent \centering{}\caption{\label{fig:mu-nu}Upper panel: Re-plotting Fig.\,\ref{fig:mu-zeta}
in terms of $\{\mu,\mu\,\zeta\}$. The dotted yellow line in Fig.\,\ref{fig:mu-zeta}
becomes a straight line $\mu+\mu\zeta\!=\!1$.\lb Lower panel: Distribution
of the sum $\mu+\mu\,\zeta$ as constructed from the upper panel;
the distribution exhibits a pronounced symmetric peak at $\mu+\mu\,\zeta=1$
and width $\sim0.2$.}
\vspace{-.3cm}
\end{figure}

The best-fit VSL--EdS model also yields $\delta=0.069$, corresponding
to $F(z)\rightarrow0.931$ at high redshift. As discussed in Ref.
\citep{VSL2024-Pantheon}, this value produces approximately a 7\%
reduction in the inferred Hubble parameter at $z\gg1$, thereby potentially
alleviating the Hubble tension. Moreover, the corresponding value
$H_{0}=47.22$ km/s/Mpc is in striking agreement with the value $H_{0}\approx46$
km/s/Mpc obtained by Blanchard, Douspis, Rowan-Robinson, and Sarkar
from their CMB fit within the EdS cosmology, likewise without invoking
dark energy \citep{BDRS-2003}.\vs\vs

\textbf{\emph{Discussion 1: Posterior alignment with the identity
$\boldsymbol{(1+\zeta)\,\mu=1}$}}---Remarkably, Fig. \ref{fig:mu-zeta}
exhibits a strong degeneracy along the locus $(1+\zeta)\,\mu=1$,
represented by the dotted yellow line in the $\left\{ \mu,\,\zeta\right\} $
plane. There is no \emph{a priori} theoretical reason for the Pantheon
data to obey this identity, which is equivalent to requiring that
$\eta\!=\!1$ (or $\gamma\!=\!1$). Thus, this \emph{posterior} alignment
is an \emph{empirical fact}. Furthermore, we find that along the locus,
for $0\leqslant\zeta\lesssim8$, the VSL power-law cosmology outperforms
the $\Lambda$CDM model in terms of $\chi^{2}$.\vs

The upper panel of Fig. \ref{fig:mu-nu} re-plots Fig. \ref{fig:mu-zeta}
using $\left\{ \mu,\,\mu\,\zeta\right\} $, where the locus transforms
into a straight line $\mu+\mu\,\zeta\!=\!1$. Given the constraints
$\left\{ \mu\!>\!0,\,\zeta\!\geqslant\!0\right\} $, the lower panel
of Fig. \ref{fig:mu-nu} shows the distribution of $\eta$ $:=(1+\zeta)\,\mu$;
it exhibits a prominent peak at $\eta\!=\!1$ with a width $\sim0.2$,
providing strong support for the relation $(1+\zeta)\,\mu=1$.\vs\vs

\textbf{\emph{Discussion 2: Relationship between speed of light and
cosmic expansion rate}}---It is straightforward to show that the
identity $(1+\zeta)\,\mu=1$, combined with Eqs. \eqref{eq:EdS-growth}
and \eqref{eq:VSL-c-general}, renders $\dot{a}=\frac{\mu}{t_{0}}(t/t_{0})^{\,\mu-1}$
and $c=c_{0}(t/t_{0})^{-\mu\zeta}=c_{0}(t/t_{0})^{\,\mu-1}$. With
$t_{0}=\mu/H_{0}$, this yields Relation \eqref{eq:key-relation},
or equivalently
\begin{equation}
c=c_{0}\,H_{0}^{-1}\,\dot{a}\ \ \ \ \ \forall t\label{eq:key-relation_2}
\end{equation}
This Relation indicates that, during the late-time epoch,\emph{ the
cosmic expansion rate is always in synchrony with the speed of light}.
This remarkable feature is starkly absent in\lb the $\Lambda$CDM
model and thus embodies a new empirical ``law''.\vs

\begin{figure}[!t]
\begin{centering}
\vskip-3pt\hskip-16pt\includegraphics[width=7cm,height=4.8cm]{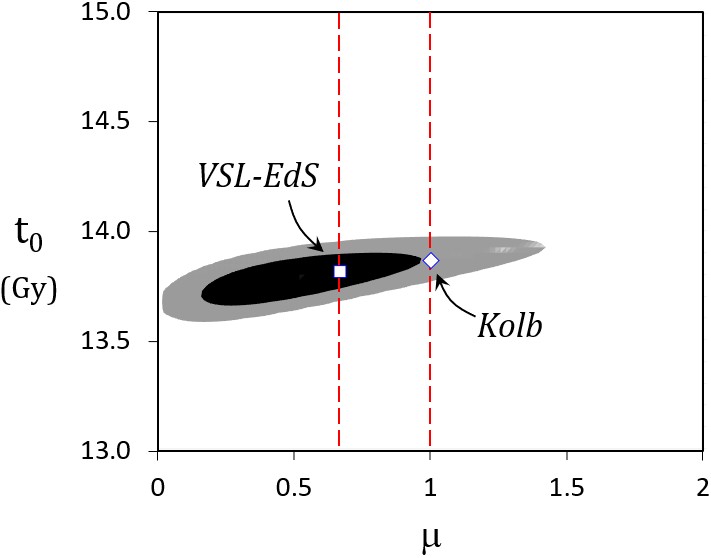}\vskip-4pt
\par\end{centering}
\caption{68\% and 95\% CL contours for models satisfying $(1+\zeta)\,\mu\!=\!1$.
The distribution of $t_{0}$ is insensitive to $\mu$. The parameters
for the VSL--EdS and Kolb cases are given in Table\,I.\lb}

\noindent \label{fig:t0-mu}\vspace{-.8cm}
\end{figure}

The VSL--EdS case, where $a\propto t^{\,2/3}$ and $c\propto a^{-1/2}\propto t^{-1/3}$,
conforms to $c\propto\dot{a}$. The Kolb case, where $a\propto t$
with $c$ held fixed, trivially satisfies this relation. Fig. \ref{fig:t0-mu}
displays the joint distribution of $\mu$ and the cosmic age $t_{0}$
for the class of models that satisfy $(1+\zeta)\,\mu=1$.\vs\vs

\textbf{\emph{Discussion 3: Generalized Copernican Principle in the
time domain}}---On a large scale, the cosmos is homogeneous and isotropic.
The RW metric embeds this Copernican Principle by allowing the scale
factor to be a function of time $t$ but not the spatial coordinates
$\{r,\theta,\varphi\}$.\vs

If $c\!=c_{0}H_{0}^{-1}\dot{a}=c_{0}H_{0}^{-1}\,aH$, the Hubble sphere
$D_{H}\!=\!c/H\!\propto a$ grows in sync with the scale factor $a$.
On a large scale, at any given instant, the recession speed between
two galaxies is synchronous with the prevailing speed of light. Consequently,
cosmic expansion is \emph{self-invariant} when viewed from any vantage
point in time.\vs

More broadly, \emph{irrespective} of the actual forms of $a(t)$ and
$c(a)$, as long as $\dot{a}\!=\!B\,c$ (where $B$ is a constant),
the Ricci scalar of the VSL--RW metric \eqref{eq:modified-RW} is
strictly proportional to $a^{-2}$, viz.\ $\mathcal{R}\!=\!6B^{2}a^{-2}$,
and its Riemann tensor is independent of $a(t)$ and $t$, viz.\ 
$\mathcal{R}_{r\theta r}^{\theta}\,$=$\,\mathcal{R}_{r\varphi r}^{\varphi}\,$=$\,-B^{2}$;
$\mathcal{R}_{\theta\theta r}^{r}\,$=$\,-\mathcal{R}_{\theta\varphi\theta}^{\varphi}\,$=$\,B^{2}\,r^{2}$;
$\mathcal{R}_{\varphi\varphi r}^{r}\,$=$\,\mathcal{R}_{\varphi\varphi\theta}^{\theta}\,$=$\,B^{2}\,r^{2}\sin^{2}\theta$
\citep{_comm-Riemann}.\lb It can be shown that the reverse statement
is also valid.\lb This property constitutes a mathematical formulation
of the Generalized Copernican Principle in the time domain.\vs\vs

\textbf{\emph{Discussion 4: An alternative explanation for late-time
acceleration}}---Along the locus $(1+\zeta)\,\mu\!=\!1$, viz. $\eta=1$,
Eq. \eqref{eq:VSL-d-vs-z} reduces to
\begin{equation}
d_{L}^{MW}=c_{MW}\,t_{0}\,\frac{1+z}{F(z)}\,\ln\frac{1+z}{F^{1+\zeta}(z)}\vspace{-.1cm}\label{eq:VSL-d-vs-z-simplified}
\end{equation}
which behaves asymptotically as
\begin{equation}
d_{L}^{MW}\,\simeq\ z\,\ln z\ \ \ \ \text{at high }z\vspace{-.1cm}\label{eq:asymptotic}
\end{equation}
Compared with the EdS universe, where\vspace{-.1cm}
\begin{equation}
d_{L}^{(EdS)}=2\,c\,\frac{1+z}{H_{0}}\,\Bigl(1-\frac{1}{\sqrt{1+z}}\Bigr)\,\simeq\ z\ \ \ \ \text{at high }z\vspace{-.1cm}\label{eq:EdS-formula}
\end{equation}
the \emph{extra factor $\ln z$} on the right-hand-side of Eq.\,\eqref{eq:asymptotic}
is responsible for the excess in distance modulus observed at high
redshift in the Hubble diagram of SNeIa. \emph{This behavior explains
why supernovae at high redshift appear dimmer than expected based
on the EdS model.} Consequently, late-time acceleration can naturally
arise from the synchronous behavior between the speed of light and
the cosmic expansion rate, without resorting to DE.\vs

The particular VSL--EdS case, where $a\!\propto\!t^{\,2/3}$ and
$c\!\propto\!a^{-1/2}$, was investigated in Ref.\ \citep{VSL2024-Pantheon}.
In that work, we showed that \emph{the phenomenon of late-time acceleration
arises from a declining speed of light in an expanding EdS universe}.
\vs\vs

\textbf{\emph{Discussion 5: Infinite cosmological and event horizons}}---If
we extend the identity $(1+\zeta)\,\mu\!=\!1$ to earlier epochs,
then the cosmological horizon is divergent:
\begin{equation}
l_{H}(t)=a(t)\!\int_{0}^{t}\frac{c(a)d\tau}{a(\tau)}\simeq\!\int_{0}^{t}\!\!\frac{d\tau}{a^{1+\zeta}(\tau)}\simeq\!\int_{0}^{t}\frac{d\tau}{\tau}=\infty
\end{equation}
This property would resolve the horizon paradox, without invoking
inflation (see also \citep{Moffat-1993,Albrecht-1998}). Furthermore,
the event horizon, defined to be the size of the region from which
a signal emitted at the current moment will ever reach the observer
(staying fixed at $\vec{x}=0$) in arbitrary distant \emph{future},
is also divergent:\vspace{-.2cm}
\begin{equation}
l_{\text{event}}(t)=a(t)\!\int_{t}^{\infty}\frac{c(a)d\tau}{a(\tau)}\simeq\int_{t}^{\infty}\frac{d\tau}{\tau}=\infty
\end{equation}

\textbf{\emph{Discussion 6: A new conformally--flat metric for cosmology}}---If
$c\,=c_{0}\,H_{0}^{-1}\,\dot{a}$, by rescaling $a\!\rightarrow\!H_{0}c_{0}^{-1}\,a$
and $r\!\rightarrow\!c_{0}H_{0}^{-1}\,r$, the (spatially-flat) VSL--RW
metric \eqref{eq:modified-RW} is transformed into \citep{_comm-conf}\vspace{-.4cm}

\begin{equation}
ds^{2}=da^{2}-a^{2}\Bigl(dr^{2}+r^{2}d\Omega_{(2)}^{2}\Bigr).\vspace{-.1cm}\label{eq:my-metric-a}
\end{equation}
In this form, the cosmic scale factor itself effectively acts as the
time coordinate; viz., each cosmic event can be labeled by the quartet
$\{a,r,\theta,\varphi\}$. Furthermore, by setting $a=e^{\,\eta}$,
Metric \eqref{eq:my-metric-a} is \emph{conformal to the Minkowski
metric}, viz.
\begin{equation}
ds^{2}=e^{\,2\eta}\,\left[d\eta^{2}-\Bigl(dr^{2}+r^{2}d\Omega_{(2)}^{2}\Bigr)\right]\label{eq:my-metric-conformal}
\end{equation}
with the conformal time coordinate $\eta\in\mathbb{R}$ \emph{unbounded}.
This conformally flat metric explains the absence of both cosmological
and event horizons, as shown in Discussion 5.\vs\vs

We \emph{conjecture} that Metric \eqref{eq:my-metric-a} (or equivalently,
\eqref{eq:my-metric-conformal}), captures the kinematics of the cosmic
scale factor. Two remarks are in order: (i) Metric \eqref{eq:my-metric-a}
is \emph{not} the Milne universe, which is spatially open and equivalent
to the Minkowski metric. In contrast, Metric \eqref{eq:my-metric-a}
is spatially flat and conformal to the Minkowski metric. (ii) Compared
with the de Sitter universe $ds^{2}\!=\!d\eta^{2}-e^{2\sqrt{\frac{\Lambda}{3}}\eta}\!\left(dr^{2}\!+\!r^{2}d\Omega^{2}\right)$
where $\eta\in\mathbb{R}$ unbounded, Metric \eqref{eq:my-metric-conformal}
has the built-in property that $c\propto\frac{da}{d\eta}\ \ \forall\eta\in\mathbb{R}$.\vs\vs\vs\vs

\textbf{\emph{Conclusion: A Novel Kinematics for Cosmology}}---Figures
\ref{fig:mu-zeta} and \ref{fig:mu-nu} capture the core finding of
our paper. Rather than imposing a specific cosmological model \emph{a
priori}, we inferred the kinematic functions $a(t)$ and $c(a)$ of
the VSL--RW metric \eqref{eq:modified-RW} directly from the Hubble
diagram of SNeIa within the scale-free Dolgov--Barrow framework.\vs\vs

Remarkably, in our approach to the Hubble diagram of SNeIa, the contour
plots conform to the relation $(1+\zeta)\,\mu=1$, as evidenced in
the lower panel of Fig.\,\ref{fig:mu-nu}. It is essential to emphasize
that\emph{ there is no a priori theoretical reason why this relation
should hold}; yet observational data support it \emph{a posteriori}
as an empirical fact.\vs\vs 

The empirical relation $(1+\zeta)\,\mu\!=\!1$ has far-reaching consequences:
it indicates that \emph{the speed of light $c$ is strictly proportional
to the cosmic expansion rate $da/dt$ at all instants during the late-time
epoch}---a striking feature that is absent in the $\Lambda$CDM model.
Consequently, as discussed in Discussions 4 and 5, the exact kinematic
relation $c=c_{0}\,H_{0}^{-1}\,\dot{a}$ naturally explains late-time
acceleration and produces an infinite cosmological horizon, all without
necessitating the inclusion of DE and inflation.\vs\vs

Although derived here within the Dolgov--Barrow cosmology, \emph{the
relation $c=c_{0}\,H_{0}^{-1}\,\dot{a}$ itself is purely kinematic.
It intrinsically relates two quantities of identical physical dimension}---the
speed of light and the cosmic expansion rate---and therefore reflects
a more general property of late-time cosmic evolution beyond the Dolgov--Barrow
parameterization.\vs\vs

If this newfound \emph{synchronous relation between $c$ and $\dot{a}$}
is further corroborated by independent probes, including the Baryon
Acoustic Oscillations and the Cosmic Microwave Background, it would
constitute a stringent \emph{kinematic} constraint that any successful
\emph{dynamical} model of cosmology must satisfy.\vspace{0.5cm}

\textbf{\emph{Acknowledgments}}---The author thanks Tiberiu Harko,
Demosthenes Kazanas, Alexander Dolgov, Soebur Razzaque, Lehel Csillag,
Pradyumn K. Sahoo, Leandros Perivolaropoulos, David Benisty, Bruce
Bassett, and Rodrigo Cuzinatto for helpful discussions.\newpage

\appendix

\section{\label{sec:Power-law-derivation}\ Derivation of the power-law form}

\emph{Statement:} Let $Y\!\!:(0,\infty)\to(0,\infty)$ be a differentiable
function satisfying
\begin{equation}
\frac{Y(x)}{Y(x_{0})}=f\!\left(\frac{x}{x_{0}}\right)\label{eq:my-form}
\end{equation}
where $f\in C^{1}(0,\infty)$. Then there exists a constant $A>0$
and an exponent $\zeta\in\mathbb{R}$ such that
\begin{equation}
Y(x)=A\,x^{\,\zeta}.
\end{equation}

\emph{Proof:} Differentiating Eq. \eqref{eq:my-form} with respect
to $x$, while holding $x_{0}$ fixed gives
\begin{equation}
\frac{1}{Y(x_{0})}\frac{dY(x)}{dx}=\frac{1}{x_{0}}f'\left(\frac{x}{x_{0}}\right).
\end{equation}
Defining $u:=\frac{x}{x_{0}}$, we obtain
\begin{equation}
\frac{d\ln Y(x)}{d\ln x}=\frac{uf'(u)}{f(u)}.
\end{equation}
The left-hand side is independent of $u$, whereas the right-hand
side is independent of $x$. Therefore, both sides must equal the
same constant $\zeta$ in $\mathbb{R}$. Using $f(1)=1$,
\begin{equation}
\frac{uf'(u)}{f(u)}=\zeta\qquad\Longrightarrow\qquad f(u)=u^{\,\zeta}.
\end{equation}
Consequently,
\begin{equation}
Y(x)=A\,x^{\,\zeta},\ \ \ A>0.
\end{equation}
QED.\vskip6pt

\emph{Justification:} Since $x$ and $Y(x)$ are \emph{dimensionful}
physical quantities, any scale-independent relation between them can
depend only on the \emph{dimensionless} ratios $x/x_{0}$ and $Y(x)/Y(x_{0})$.

\section{\label{sec:Freq-ratio}\ Frequency ratio in VSL cosmology}

The null geodesic ($ds^{2}=0$) for a lightwave traveling from an
emitter toward Earth (viz. $d\Omega_{(2)}=0$) is thus: $dr=\frac{c_{0}\,dt}{a^{1+\zeta}(t)}$.
Denote $t_{em}$ and $t_{ob}$ the emission and observation time points
of the lightwave, and $r_{em}$ the co-moving distance of the host
galaxy from the Milky Way. From \eqref{eq:RW}, we have\vspace{-.3cm}
\begin{equation}
r_{em}=c_{0}\int_{t_{em}}^{t_{ob}}\!\!\frac{dt}{a^{1+\zeta}(t)}
\end{equation}
The next wavecrest to leave the emitter at $t_{em}+\delta t_{em}$
and arrive at Earth at $t_{ob}+\delta t_{ob}$ satisfies\vspace{-.1cm}
\begin{equation}
r_{em}=c_{0}\int_{t_{em}+\delta t_{em}}^{t_{ob}+\delta t_{ob}}\!\!\frac{dt}{a^{1+\zeta}(t)}
\end{equation}
Subtracting the two equations yields
\begin{equation}
\frac{\delta t_{ob}}{a_{ob}^{1+\zeta}}=\frac{\delta t_{em}}{a_{em}^{1+\zeta}}
\end{equation}
which leads to the ratio between the emitted frequency and the observed
frequency\vspace{-.1cm}
\begin{equation}
\frac{\nu_{ob}}{\nu_{em}}=\frac{\delta t_{em}}{\delta t_{ob}}=\frac{a_{em}^{1+\zeta}}{a_{ob}^{1+\zeta}}=a^{1+\zeta}.
\end{equation}

\section{\label{sec:Modified-d-z}\ Modifying the distance--redshift formulae
for VSL cosmology}

\noindent Note that with $a(t)=\left(t/t_{0}\right)^{\mu}$, we obtain
(by defining $\eta:=(1+\zeta)\,\mu$)\vspace{-.1cm}
\begin{align}
r & =c_{0}\int_{t}^{t_{0}}\frac{dt}{a^{1+\zeta}(t)}=c_{0}\int_{t}^{t_{0}}\frac{dt}{\left(t/t_{0}\right)^{\mu(1+\zeta)}}\\
 & =\frac{c_{0}t_{0}}{1-\eta}\left[1-\left(\frac{t}{t_{0}}\right)^{-\eta+1}\right]=\frac{c_{0}t_{0}}{1-\eta}\left[1-a^{\frac{-\eta+1}{\mu}}\right]
\end{align}
Using $1+z=a^{-(1+\zeta)}F^{1+\zeta}(z)$ , we obtain the \emph{modified}
distance--redshift formula\vspace{-.1cm}
\begin{align}
r & =\frac{c_{0}\,t_{0}}{1-\eta}\!\left[1-\left(\frac{1+z}{F^{1+\zeta}(z)}\right)^{1-\frac{1}{\eta}}\right]
\end{align}

\noindent As in standard cosmology, the luminosity distance $d_{L}$
is defined via the absolute luminosity $L$ and the apparent luminosity
$J$ by\vspace{-.3cm}
\begin{equation}
d_{L}^{\,2}=\frac{L}{4\pi J}\label{eq:definition-of-dL}
\end{equation}
On the other hand, the absolute luminosity $L$ and the apparent luminosity
$J$ are related\vspace{-.1cm}
\begin{equation}
4\pi r^{2}J=L\,\frac{\lambda_{SN}}{\lambda_{MW}}.\frac{\lambda_{SN}}{\lambda_{MW}}\label{eq:J-vs-L}
\end{equation}
In the RHS of \eqref{eq:J-vs-L}, the first term $\lambda_{SN}/\lambda_{MW}$
represents the ``loss'' in the energy of the red-shifted photon known
as the ``Doppler theft''. The second (identical) term $\lambda_{SN}/\lambda_{MW}$
is due to the dilution factor in the photon density as the same number
of photons get distributed in the altered wavecrest in the radial
direction (i.e., the light ray). The $4\pi r^{2}$ in the LHS of \eqref{eq:J-vs-L}
is the spherical dilution in flat space. Combining Eqs. \eqref{eq:definition-of-dL},
\eqref{eq:J-vs-L}, \eqref{eq:tmp} and the definition of $F$ (that
follows \eqref{eq:tmp}), we get\vspace{-.1cm}
\begin{align}
d_{L} & =r\,\frac{\lambda_{MW}}{\lambda_{SN}}=r\,\frac{\lambda_{MW}}{\lambda_{MW}^{\star}}.\frac{\lambda_{MW}^{\star}}{\lambda_{SN}}=r\,(1+z)\,\frac{1}{F(z)}
\end{align}

\noindent Due to the refraction effect at Transit \#3, the apparent
luminosity distance observed by the Earth-based astronomer $d_{L}^{MW}$
differs from $d_{L}$ by the factor $c_{MW}/c_{0}$, viz. $\frac{d_{L}^{MW}}{c_{MW}}=\frac{d_{L}}{c_{0}}$.
Finally, we obtain the \emph{modified }luminosity distance--redshift
relation
\begin{align}
d_{L}^{MW} & =\frac{c_{MW}}{c_{0}}.\frac{\lambda_{MW}}{\lambda_{SN}}\,r\\
 & =\frac{c_{MW}\,t_{0}}{1-\eta}\,\frac{1+z}{F(z)}\left[1-\left(\frac{1+z}{F^{1+\zeta}(z)}\right)^{1-\frac{1}{\eta}}\right].
\end{align}


\begin{thebibliography}{test}
\bibitem{_comm-Riemann}More generally, consider a VSL--RW metric
$ds^{2}=c^{2}(a)dt^{2}-a^{2}(t)\bigl(\frac{dr^{2}}{1-\varkappa r^{2}}+r^{2}d\Omega^{2}\bigr)$
with $\varkappa\in\left\{ -1,0,1\right\} $. When $\dot{a}=B\,c$,
its Ricci scalar $\mathcal{R}=6\,\frac{\varkappa+B^{2}}{a^{2}}$ is
strictly proportional to $a^{-2}$, and its Riemann tensor components
$\mathcal{R}_{r\theta r}^{\theta}\,$=$\,\mathcal{R}_{r\varphi r}^{\varphi}\,$=$\,-\frac{\varkappa+B^{2}}{1-\varkappa r^{2}}$;
$\mathcal{R}_{\theta\theta r}^{r}\,$=$\,-\mathcal{R}_{\theta\varphi\theta}^{\varphi}\,$=$\,\left(\varkappa+B^{2}\right)\,r^{2}$;
$\mathcal{R}_{\varphi\varphi r}^{r}\,$=$\,\mathcal{R}_{\varphi\varphi\theta}^{\theta}\,$=$\,\left(\varkappa+B^{2}\right)\,r^{2}\sin^{2}\theta$
are independent of $a(t)$ and $t$.
\bibitem{_comm-conf}For a general spatial curvature, i.e. with $\varkappa\in\left\{ -1,0,1\right\} $,
Metric \eqref{eq:my-metric-a} reads $ds^{2}=da^{2}-a^{2}\bigl(\frac{dr^{2}}{1-\varkappa r^{2}}+r^{2}d\Omega^{2}\bigr)$
and Metric \eqref{eq:my-metric-conformal} is $ds^{2}=e^{\,2\eta}\,\bigl[d\eta^{2}-\bigl(\frac{dr^{2}}{1-\varkappa r^{2}}+r^{2}d\Omega^{2}\bigr)\bigr]$.
\end{thebibliography}
\end{document}